\documentclass[twocolumn,preprintnumbers,amsmath,amssymb]{revtex4}

\usepackage{graphicx}
\usepackage{dcolumn}
\usepackage{bm}

\begin{document}

\title{Critical opalescence in fluids: 1.5-Scattering effects and \\the Landau--Placzek ratio}

\author{M. Ya.~Sushko}
\altaffiliation [\textit{E-mail address}: ]{mrs@onu.edu.ua (M. Ya.
Sushko)} \affiliation {Department of Theoretical Physics, Odessa
I. I. Mechnikov
National University, \\
2 Dvoryanska St., Odessa 65026, Ukraine}

\begin{abstract}
\noindent \textbf{Abstract}

We adduce new arguments for the significance of so-called 1.5- (or
sesquialteral) molecular light scattering in one-component fluids.
For this purpose, we analyze its effect on the Landau--Placzek
ratio for the critical opalescence spectrum. The results obtained
are used to reveal experimental data which can be interpreted as
evidence for its existence and to evaluate both the relative
magnitude and the sign of the 1.5-scattering contribution.

\bigskip


\end{abstract}

\maketitle

\bigskip

\section{Introduction}
The accuracy of experimental data obtained with the method of
molecular spectroscopy for the thermodynamic and kinetic
parameters of fluids depends heavily on the validity of
theoretical views of the physical mechanisms responsible for the
formation of the spectra under one or other circumstances. In
particular, it is widely believed that the critical opalescence
spectra from one-component fluids are formed only by single
\cite{1,2,3,4,5} and true multiple scattering effects
\cite{6,7,8,9,10,11,12}. The latter are treated as successive
single scatterings by distant (satisfying the wave-zone condition
$r > > \lambda $, where $\lambda $ is the wavelength of probing
radiation in the fluid) pairs of density fluctuations.

Recently, we suggested \cite{13,14,15} that besides the above
mechanisms, a so-called 1.5- (or sesquialteral) scattering can
become of significance near the critical point. This scattering is
formed by those three-point configurations of density fluctuations
in which the fluctuations are spaced by distances $r < < r_{c} <
\lambda $, $r_{c} $ being the correlation radius. Groups of
fluctuations satisfying this condition are termed further as
compact.

Based on theory \cite{13,14,15}, we have managed to estimate the
intensity and the sign of 1.5-scattering, identify the
thermodynamic paths most favorable for its observation, and
analyze its effect on the depolarization ratio and the Rayleigh
line halfwidth of the critical opalescence spectra. The
1.5-scattering contribution proves to be most pronounced at a
certain distance from the critical point. Asymptotically close to
the critical point, it is expected to vanish, in line with
Polyakov's hypothesis \cite{16,17} of conformal symmetry of
critical fluctuations.

Here, we give a brief account of our analysis of the
1.5-scattering effect on the Landau--Placzek ratio for the
critical opalescence spectra. The results obtained enable us to
quote new experimental data in support of the existence of the
1.5-scattering contribution and to make new independent estimates
of its relative magnitude and sign.

\section {General structure of the single-scattering spectrum}

According to Refs. \cite{13,15}, the spectrum of polarized
single-scattering is given by
\begin{equation}
\label{eq1}
I({\rm {\bf q}},\omega ) = {\sum\limits_{n,m = 1} {I_{nm} ({\rm {\bf
q}},\Omega )}} ,
\end{equation}
\noindent where the term
\begin{eqnarray}
I_{nm} ({\rm {\bf q}},\Omega ) \propto \,\left( { -
{\frac{{1}}{{3\varepsilon _{0}}} }} \right)^{n + m - 2} \nonumber
\\ \times {\frac{{1}}{{\pi }}}\text{Re}{\int\limits_{0}^{ + \infty} {dt}}
{\int\limits_{V} {d{\rm {\bf r}}{\left\langle {\left( {\delta
\varepsilon ({\rm {\bf r}},t)} \right)^{n}\left( {\delta
\varepsilon (0,0)} \right)^{m}} \right\rangle }\,e^{i\Omega t -
i{\rm {\bf q}} \cdot {\rm {\bf r}}}}} , \label{eq2}
\end{eqnarray}
\noindent represents the contribution from a pair of compact
groups of $n$ and $m$ permittivity fluctuations, and $\Omega $ and
${\rm {\bf q}}$ are the changes in the frequency and wavevector of
the wave due to scattering. Expression (\ref{eq2}) is the
frequency transform of the function
\begin{eqnarray}
I_{nm} ({\rm {\bf q}},t) \propto \,\left( { -
{\frac{{1}}{{3\varepsilon _{0} }}}} \right)^{n + m - 2} \nonumber \\
\times {\int\limits_{V} {d{\rm {\bf r}}{\left\langle {\left(
{\delta \varepsilon ({\rm {\bf r}},t)} \right)^{n}\left( {\delta
\varepsilon (0,0)} \right)^{m}} \right\rangle} e^{ - i{\rm {\bf
q}} \cdot {\rm {\bf r}}}}} ,\label{eq3}
\end{eqnarray}
\noindent whose value at $t = 0$ is equal to the integrated
intensity of scattering by the above pair.

We accept further that for one-component fluids, the permittivity
fluctuations are mainly due to the density ($\rho )$ fluctuations $\delta
\rho ({\rm {\bf r}},t)$: $\delta \varepsilon ({\rm {\bf r}},t) \approx
\left( {{{\partial \varepsilon}  \mathord{\left/ {\vphantom {{\partial
\varepsilon}  {\partial \rho}} } \right. \kern-\nulldelimiterspace}
{\partial \rho}} } \right)\delta \rho ({\rm {\bf r}},t)$.

\section {Spectrum of the ``standard'' single scattering}

It is only the term $I_{11} (q,\Omega )$ in Eq. (\ref{eq1}) that
is customarily believed to form (besides the true multiple
scattering) the critical opalescence spectra. With sufficient
accuracy, it can be analyzed under the assumptions that both the
linearized hydrodynamic equations and the concept of local
thermodynamic equilibrium are valid. For the case where the
transverse part of the velocity is not coupled with $\delta \rho
({\rm {\bf r}},t)$, the lowest-order term of the Fourier component
$\rho _{{\rm {\bf k}}} (t)$ of $\delta \rho ({\rm {\bf r}},t)$
evolves in time by the law \cite{1,4}
\begin{equation}
\label{eq4}
\rho _{{\rm {\bf k}}} \left( {t} \right) = \rho _{{\rm {\bf k}}} \left( {0}
\right){\left[ {{\frac{{c_{P} - c_{V}}} {{c_{P}}} }e^{ - \Gamma _{{\rm R}}
(k)t} + {\frac{{c_{V}}} {{c_{P}}} }e^{ - \Gamma _{{\rm B}} (k)\,t}\cos
\left( {u_{S} kt} \right)} \right]},
\end{equation}
\noindent where $$\Gamma _{{\rm R}} (k) = {\frac{{\kappa}} {{\rho
c_{P}}} }k^{2}, \Gamma _{{\rm B}} (k) = {\frac{{1}}{{2\rho}}
}{\left[ {{\frac{{4}}{{3}}}\eta + \zeta + \kappa \left(
{{\frac{{1}}{{c_{V}}} } - {\frac{{1}}{{c_{P}}} }} \right)}
\right]}\,k^{2},$$ $c_{P} $ and $c_{V} $ are the specific heats at
constant pressure and volume ($c_{P} / c_{V} \equiv \gamma )$,
$\kappa $ is the thermal conductivity, $\eta $ and $\zeta $ are
the shear and bulk viscosities, and $u_{S} $ is the adiabatic
speed of sound. Correspondingly,
\begin{equation}
\label{eq5}
I_{11} (q,t) \propto \left( {{\frac{{\partial \varepsilon}} {{\partial \rho
}}}} \right)^{2}G(q){\left[ {A\,e^{ - \Gamma _{{\rm R}} (q)t} + B\,e^{ -
\Gamma _{{\rm B}} (q)\,t}\cos (u_{S} qt)} \right]},
\end{equation}
\noindent where $G(q) \equiv {\left\langle {\rho _{{\rm {\bf q}}}
(0)\rho _{ - {\rm {\bf q}}} (0)} \right\rangle} $, $A \equiv
(c_{P} - c_{V} ) / c_{P} $, and $B \equiv c_{V} / c_{P} $.

The two addends in the brackets (\ref{eq5}) signify that both
thermal and mechanical processes contribute to the evolution of
$\delta \rho ({\rm {\bf r}},t)$. They lead to the formation of
three lines in the spectrum of the ``standard'' single scattering,
with total integrated intensity $I_{11} (q) \propto \left(
{{{\partial \varepsilon}  \mathord{\left/ {\vphantom {{\partial
\varepsilon}  {\partial \rho}} } \right.
\kern-\nulldelimiterspace} {\partial \rho}} } \right)^{2}G(q)$.
One of these, not shifted in frequency, is the Rayleigh line, with
halfwidth $\Gamma _{{\rm 1}{\rm R}} = \Gamma _{{\rm R}} (q)$ and
integrated intensity $I_{{\rm 1}{\rm R}} = AI_{11} (q)$, and the
others, shifted, are the Brillouin lines, with halfwidth $\Gamma
_{{\rm 1}{\rm B}} = \Gamma _{{\rm B}} (q)$ each and total
integrated intensity $I_{{\rm 1}{\rm B}} = BI_{11} (q)$. The
quotient
\begin{equation}
\label{eq6}
R = {\frac{{I_{{\rm 1}{\rm R}}}} {{I_{{\rm 1}{\rm B}}}} } =
{\frac{{A}}{{B}}} = \gamma - 1.
\end{equation}
\noindent is known as the Landau--Placzek ratio.

As the critical point is approached, long-range density
correlations become of crucial importance and, strictly speaking,
the applicability of the above approach becomes questionable. Yet
the structure of Eq. (\ref{eq4}) is usually preserved by
suggesting that the kinetic coefficients can be treated as
functions of temperature, wave vector and, in general, frequency.
Some of the approaches offered to recover the explicit forms of
these functions are reviewed in Refs. \cite{5,12}.

\section {Spectrum of the 1.5-scattering}

The 1.5-scattering spectrum is represented in Eq. (\ref{eq1}) by the terms with $n =
1$, $m = 2$ and $n = 2$, $m = 1$. For one-component fluids, its inverse
Fourier transform is
\begin{eqnarray}
I_{1.5} (q,t) \propto - {\frac{{1}}{{3\varepsilon _{0}}} }\left(
{{\frac{{\partial \varepsilon}} {{\partial \rho}} }} \right)^{3}
\int\limits_{V} d{\rm {\bf r}}\,e^{i{\rm {\bf q}} \cdot {\rm {\bf
r}}} \nonumber \\\times {\left[ {{\left\langle {\delta \rho ({\rm
{\bf r}},t)(\delta \rho ({\rm {\bf 0}},0))^{2}} \right\rangle} +
{\left\langle {(\delta \rho ({\rm {\bf r}},t))^{2}\delta \rho
({\rm {\bf 0}},0)} \right\rangle}}  \right]} . \label{eq7}
\end{eqnarray}

To evaluate the effect of 1.5--scattering on the overall spectrum,
we follow the following procedure. First, we assume in the
standard way \cite{1,4} that the time evolution of $\rho _{{\rm
{\bf k}}} (t)$ is given by Eq. (\ref{eq4}). Then we change to the
wave-vector space to represent the fluctuation $\delta \rho ({\rm
{\bf r}},t)$ as
\begin{equation}
\label{eq8}
\delta \rho ({\rm {\bf r}},t) = {\sum\limits_{{\rm {\bf k}}} {\rho _{{\rm
{\bf k}}} (0)}} {\left[ {Ae^{ - \Gamma _{{\rm R}} (k)t} + Be^{ - \Gamma
_{{\rm B}} (k)\,t}\cos (u_{S} kt)} \right]}\,e^{i{\rm {\bf k}}{\rm {\bf
r}}},
\end{equation}
\noindent where we consider the quantities $A$ and $B$ as
$k$--independent. With representation (\ref{eq8}), the statistical
averaging in Eq. (\ref{eq7}) is equivalent to averaging over all
realizations of the initial values $\rho _{{\rm {\bf k}}} \left(
{0} \right) \equiv \,\rho _{{\rm {\bf k}}} $. In doing so, we take
advantage of the approximation \cite{15} (${c}'$ is a parameter)
\begin{equation}
\label{eq9}
{\left\langle {\rho _{{\rm {\bf k}}_{1}}  \rho _{{\rm {\bf k}}_{12}}  \rho
_{{\rm {\bf k}}_{3}}}   \right\rangle}  \approx - {\frac{{2{c}'}}{{k_{B}
T\sqrt {V}}} }G(k_{1} )G(k_{2} )G(k_{3} )\,\delta _{{\rm {\bf k}}_{3} , -
{\rm {\bf k}}_{1} - {\rm {\bf k}}_{2}}
\end{equation}
\noindent for the three-point correlation function of density
fluctuations. This gives:
\begin{widetext}
\begin{equation}
\label{eq10}
I_{1.5} (q,t) = {\sum\limits_{i - 1}^{5} {I_{1.5}^{(i)} (q,t)}} ,
\end{equation}
\begin{equation}
\label{eq11}
I_{1.5}^{\left( {1} \right)} (q,t) \propto Ae^{ - \Gamma _{{\rm R}}
(q)t}KG(q)\int {d{\rm {\bf k}}\,G(k)G(\vert {\rm {\bf q}} - {\rm {\bf
k}}\vert )} ,
\end{equation}
\begin{equation}
\label{eq12}
I_{1.5}^{\left( {2} \right)} (q,t) \propto Be^{ - \Gamma _{{\rm B}}
(q)t}\cos (u_{S} qt)KG(q)\int {d{\rm {\bf k}}\,G\left( {k} \right)G(\vert
{\rm {\bf q}} - {\rm {\bf k}}\vert )} ,
\end{equation}
\begin{equation}
\label{eq13}
I_{1.5}^{\left( {3} \right)} (q,t) \propto A^{2}KG(q)\int {d{\rm {\bf
k}}\,G(k)G(\vert {\rm {\bf q}} - {\rm {\bf k}}\vert )e^{ - \Gamma _{{\rm R}}
(k)t - \Gamma _{{\rm R}} (\vert {\rm {\bf q}} - {\rm {\bf k}}\vert )t}} {\rm
,}
\end{equation}
\begin{equation}
\label{eq14}
I_{1.5}^{\left( {4} \right)} (q,t) \propto 2ABKG(q)\,\int {d{\rm {\bf
k}}\,G(k)G(\vert {\rm {\bf q}} - {\rm {\bf k}}\vert )e^{ - \Gamma _{{\rm R}}
(k)t - \Gamma _{{\rm B}} (\vert {\rm {\bf q}} - {\rm {\bf k}}\vert )t}\cos
(u_{S} \vert {\rm {\bf q}} - {\rm {\bf k}}\vert t)} ,
\end{equation}
\begin{equation}
\label{eq15}
I_{1.5}^{\left( {5} \right)} (q,t) \propto B^{2}KG(q)\int {d{\rm {\bf
k}}\,G(k)G(\vert {\rm {\bf q}} - {\rm {\bf k}}\vert )e^{ - \Gamma _{{\rm B}}
(k)t - \Gamma _{{\rm B}} (\vert {\rm {\bf q}} - {\rm {\bf k}}\vert )t}\cos
(u_{S} k\,t)\cos (u_{S} \vert {\rm {\bf q}} - {\rm {\bf k}}\vert t)} {\rm
,}
\end{equation}
\end{widetext}
\noindent where $K \equiv {\frac{{2}}{{3\varepsilon _{0} (2\pi
)^{3}}}}{\frac{{c'}}{{k_{B} T}}}\left( {{\frac{{\partial
\varepsilon }}{{\partial \rho}} }} \right)^{3}$. The
1.5--scattering spectrum is given by the sum of the frequency
transforms of the functions (\ref{eq11})--(\ref{eq15}), and its
integrated intensity is ($A + B = 1)$
\begin{equation}
\label{eq16}
I_{1.5} (q) \propto 2KG(q)\int {d{\rm {\bf k}}\,G\left( {k} \right)G(\vert
{\rm {\bf q}} - {\rm {\bf k}}\vert )} .
\end{equation}

Finally, we analyze the frequency distributions of terms (\ref{eq11})--(\ref{eq15}) and
estimate their combined contributions $I_{1.5{\rm R}} $ and $I_{1.5{\rm B}}
$ to the integrated intensities of the Rayleigh and Brillouin lines
in the overall spectrum. Since the term (\ref{eq15}) contributes (in a proportion
$x$ to $1 - x)$ to both, we obtain:
\begin{equation}
\label{eq17} I_{1.5{\rm R}} \approx I_{1.5}^{(1)} (q,0) +
I_{1.5}^{(3)} (q,0) + xI_{1.5}^{(5)} (q,0) = aAI_{1.5} (q),
\end{equation}
\begin{equation}
\label{eq18} I_{{\rm 1}.{\rm 5}{\rm B}} \approx I_{1.5}^{(2)}
(q,0) + I_{1.5}^{(4)} (q,0) + (1 - x)I_{1.5}^{(5)} (q,0) =
bBI_{1.5} (q),
\end{equation}
\begin{equation}
\label{eq19}
a = {\frac{{1}}{{2}}}\left( {1 + A + x{\frac{{B^{2}}}{{A}}}} \right),
\quad
b = {\frac{{1}}{{2}}}{\left[ {1 + 2A + (1 - x)B} \right]}.
\end{equation}

\section {Effect of 1.5--scattering on the Landau--Placzek ratio}

With 1.5--scattering effects present and in view of Eqs. (\ref{eq6}) and (\ref{eq17})--(\ref{eq19}),
the Landau--Placzek ratio takes the form
\begin{equation}
\label{eq20}
R_{{\rm e}{\rm x}{\rm p}} = {\frac{{I_{{\rm 1}{\rm R}} + I_{{\rm 1}{\rm
.}{\rm 5}{\rm R}}}} {{I_{1{\rm B}} + I_{1.5{\rm B}}}} } = R{\frac{{1 +
af}}{{1 + bf}}},
\end{equation}
\noindent where the quantity $f \equiv I_{1.5} (q) / I_{11} (q)$
has the meaning of the relative magnitude of the 1.5-scattering
intensity as compared to that of the ``standard'' single
scattering. From Eq. (\ref{eq20}),
\begin{equation}
\label{eq21}
f = {\frac{{R - R_{\exp}} } {{bR_{\exp}  - aR}}}.
\end{equation}
It follows that combining spectroscopic ($R_{{\rm e}{\rm x}{\rm
p}} )$ and caloric ($c_{P} $ and $c_{V} )$ data opens, in
principle, an opportunity for direct experimental estimation of
the 1.5-scattering contribution.

\bigskip
\section {Situation with experiment}

We have found in the literature only a few measurements of the
Landau--Plazcek ratio for pure fluids near the critical point. In
Refs. \cite{18,19}, $R$ was studied for CO$_{{\rm 2}}$ along the
critical isochore above $T_{c} $ (the temperature regions were
about $3 \cdot 10^{ - 4} < \tau < 3 \cdot 10^{ - 2})$. In both
works, the $\tau $--dependence of $R$ was approximated by a single
exponent ($0.95\pm 0.15$ and $1.02\pm 0.03$ respectively). In Ref.
\cite{20}, $R$ was measured for saturated $n$-butane along the
vapor ($2.2 \cdot 10^{ - 3} < \tau < 0.08)$ and liquid ($3.3 \cdot
10^{ - 3} < \tau < 0.3)$ branches of the coexistence curve. For
$\tau < 0.1$, the experimental data were also approximated by
straight lines for both branches, while the values of $\gamma -
1$, calculated with data \cite{21}, exhibited a small curvature.
However, for the vapor the deviations were within the limits of
error, and for the liquid there was a small deviation just outside
the limit of error for $\tau < 3.5 \cdot 10^{ - 3}$.

According to our estimates \cite{13,14,15}, the conditions in the
above experiments were unfavorable for observation of
1.5--scattering. In contrast, of great interest are the results of
Ref. \cite{22} for liquid He close to the $\lambda $-line, where
the measurements of $R$ were carried out over the temperature
range $0.01\,{\rm m}{\rm K} < {\rm \vert} \Delta T{\rm \vert}  <
{\rm 1}{\rm 0}\,{\rm m}{\rm K}$ both below and above the $\lambda
$-transition. Most of the measurements were made for a pressure of
18.24 bars and some for 26 bars. Provided ${\rm \vert} \Delta
T{\rm \vert}  > {\rm 0}.{\rm 3}\,{\rm m}{\rm K}$, the observed
values of $R$ were in good agreement with the values of $\gamma -
1$, calculated with data \cite{23}, but for smaller ${\rm \vert}
\Delta T{\rm \vert} $ they were observed to fall below $\gamma -
1$.

Consider the case of He above the $\lambda $-line, where it
behaves as an ordinary liquid. According to data \cite{22,23}, for
$\Delta T \approx 0.04\,{\rm m}{\rm K}$ and a pressure of 18.24
bars $R_{{\rm e}{\rm x}{\rm p}} \approx 0.16$ and $\gamma \approx
1.21$ ($R \approx 0.21)$. Taking $x \approx 0.5$, we find $f
\approx - 0.26$. This new result agrees well with our earlier
estimates \cite{14,15}, obtained by processing the depolarization
ratio data for Xe.

I thank Professor M. A. Anisimov and Professor A. V. Chaly\u{\i}
for discussion of the results.

\bigskip
\_\_\_\_\_\_\_\_\_\_\_\_\_\_\_\_\_\_\_\_\_\_\_\_\_\_\_\_\_\_\_\_\_\_\_\_\_

This is part of the report presented at a Section III (Phase
transitions and critical phenomena) Oral Session of
 5$^{{\rm t}{\rm h}}$ International Conference \textit{Physics of Liquid Matter: Modern Problems},
  21--24 May, 2010, Kyiv, Ukraine.

\end{document}